\documentstyle[12pt]{article}
\def\be{\begin{equation}}
\def\h{\cal H}

\title{Conservation Laws and 
Cosmological Perturbations in Curved Universes}
\author{Jean-Philippe Uzan$^1$, Nathalie Deruelle$^{1,2}$ and
        Neil Turok$^2$ \\
       \and
       $^1$ D\'epartement d'Astrophysique Relativiste et de Cosmologie\\
       UPR 176 du Centre National de la Recherche Scientifique \\
       Observatoire de Paris, 92195 Meudon, France\\ 
       \and
       $^2$ Department of Applied Mathematics and Theoretical Physics\\
       University of Cambridge, Silver Street\\
       Cambridge CB3 9EW, England}

\begin{document}
\maketitle

%---------------------------------------------------------------------
\begin{abstract}
%---------------------------------------------------------------------
\vskip0.5cm
\noindent
When working in synchronous gauges, pseudo-tensor conservation laws
are often used to set the initial conditions for cosmological scalar
perturbations, when those are generated by topological defects which
suddenly appear in an up to then perfectly homogeneous and isotropic
universe. However those conservation laws are restricted to spatially
flat ($K=0$) Friedmann-Lema\^\i tre spacetimes. In this paper, we first show
that in fact they implement a matching condition between the pre-
and post- transition eras and, in doing so, we are able to generalize them
and set the initial conditions for all $K$. Finally, in the long
wavelength limit, we encode them into a vector conservation law having a well-defined
geometrical meaning.
\end{abstract}

\noindent {\bf Pacs~:} 98.80.Hw, 98.70.Vc, 04.20.Cv\\
\noindent{\bf Ref. PRD~:} DD6306
%---------------------------------------------------------------------

%---------------------------------------------------------------------
\section{Introduction}
%---------------------------------------------------------------------

Topological defects act in cosmology as ``active" sources for the
metric and matter perturbations in that the linear differential
equations for the evolution of these perturbations are inhomogeneous
(in contrast with what happens in inflationary scenarios). Thus the
general solution of the evolution equation for each perturbation mode
is the sum of a particular solution of the complete equation and the
general solution of the homogeneous equation. It depends on
integration constants that have to be fixed by the physics which gave
rise to the defects.

One has thus to take into account the fact that the defects appear at
a phase transition in a previously strictly homogeneous and isotropic
universe. For scales much larger than the Hubble radius at that time,
which are the scales of interest today, the phase transition appears
as instantaneous (in the following we shall use indifferently the words
``scales" or ``wavelengths"). A way to fix the initial conditions is
therefore to find linear combinations of the perturbation variables which
are constant in time and force them to vanish since they were strictly zero
before the phase transition. It has been shown
\cite{deruelle98} that, in the newtonian gauge, the initial conditions 
are completely fixed
by these matching conditions.

In synchronous gauges, for flat ($K=0$) Friedmann-Lema\^\i tre
spacetimes, and for ``scalar" perturbations (see \cite{bardeen80} for
the decomposition of perturbations into scalar, vector and tensor
modes), the initial conditions are usually fixed, at least partially
(see e.g.  \cite{pen94}, \cite{turok96}, \cite {hu96}, \cite{hu97})), by
means of a ``conserved pseudo-tensor" $\tau_{\mu\nu}$
\cite{pen94}, more precisely by the constraint ``$\tau_{00}=0$" (the
equation of conservation
$\partial_0\tau_{00}=\partial_k\tau_{0k}\simeq 0$ on large scales then
ensures that this constraint is conserved during the evolution
\cite{pen94}).

In this article, we first compare and contrast the implementation of
the initial conditions for long wavelength scalar perturbations in
both gauges (\& 2 and \& 3). To do so we consider a toy model, namely
coherent defects in a radiation dominated, spatially flat,
universe. We stress some of the Charybdes and Scyllas of studying its
perturbations in synchronous gauges, and show how the
``pseudo-tensor'' $\tau_{\mu\nu}$ defined in
\cite{pen94} partially fixes the initial conditions.  We then interpret 
the condition
``$\tau_{00}=0$" as a matching condition between the pre- and post-
transition eras.  Finally, we show how the newtonian gauge
simplifies the resolution of the evolution equations.  In \& 4,
because we had related it to a matching condition in the flat case, we
can construct a generalized  ``conserved pseudo-tensor" in synchronous
gauges for $K\neq0$ Friedmann-Lema\^\i tre spacetimes. Finally, in \&
5, we formalize the definition of this ``pseudo-tensor" by means of a
well defined vector conservation law \cite{katz97} and give it a precise
geometrical meaning.

%---------------------------------------------------------------------
\section{Scalar perturbations in synchronous gauges}
%---------------------------------------------------------------------

The line element of a perturbed Friedmann-Lema\^\i tre spacetime reads,
when the perturbations are scalar and in synchronous gauges~:
\be 
ds^2=a^2(\eta)\left[-d\eta^2+\left\lbrace (1+\frac{1}{3}h)\gamma_{ij}
+2(D_iD_j-\frac{1}{3}\gamma_{ij}\Delta)E\right\rbrace
dx^idx^j\right]\label{metric}
\end{equation} 
where $\eta$ is conformal time, $a(\eta)$ the background scale
factor~; $\gamma_{ij}$ is the three-metric of the constant time
maximally symmetric surfaces in the coordinate system $x^i$
($i=1,2,3)$~; $D_i$ is the associated covariant derivative,
$\Delta\equiv D_iD^i$ (latin indices are raised by means of the inverse metric
$\gamma^{ij}$)~; finally
$h$ and
$E$ are two ``small" functions of space and time.

Infinitesimal coordinate
transformations which preserve the synchronicity of the gauge can be
performed. They are defined by~:
\be
\eta\rightarrow\eta+T\qquad,\qquad x^k\rightarrow x^k+\partial^kL \label{jauge}
\end{equation}
where $T$ and $L$ are two  first
order functions of $\eta$ and $x^i$ such that
\be 
T'+{\cal H}T=0\quad\hbox{and}\quad L'-T=0\quad \Longleftrightarrow\quad
T=\frac{e_0}{a}\quad\hbox{and}\quad L=e_1+e_0\int\frac{d\eta}{a}\label{jauge2}
\end{equation} 
where a prime denotes a derivative with respect to conformal time, where
${\cal H}\equiv a'/a$ and where $e_0$ and $e_1$ are functions of space
only. The two scalar metric perturbations $h$ and $E$ then transform as
\be 
h\rightarrow h+6{\cal H}T+2\Delta L\quad ,\quad E\rightarrow E+L\label{tjauge}
\end{equation}
with $T$ and $L$ given by (3).

The energy-momentum tensor of the matter content of this perturbed
universe can be written as~: 
\be 
T_{\mu\nu}=\bar T_{\mu\nu}+\delta T_{\mu\nu}+\Theta_{\mu\nu}.\label{tmunu}
\end{equation}
$\bar T_{\mu\nu}$
($\mu,\nu=0,1,2,3$) is the energy-momentum tensor of the homogeneous
and isotropic background of radiation and dust~; $\delta T_{\mu\nu}$ is its
perturbation~: in synchronous gauges, its scalar components can be expressed
 in terms of
two scalar quantities, $\delta\equiv\delta\rho/\rho$, the density
contrast, and $v$, the velocity perturbation, as (see e.g. \cite{deruelle98})~:
\be 
\delta T_{00}=\rho a^2\delta,\quad
\delta T_{0i}=-\rho a^2(1+\omega)D_i v,\quad
\delta T_{ij}=Pa^2\left[\left(\frac{h}{3}+\Gamma+
\frac{c_s^2}{\omega}\delta\right)\gamma_{ij}+2D_iD_jE\right].\label{deltaT}
\end{equation}
$\rho$ and $P$ are the density and pressure of the background fluid,
$\omega\equiv P/\rho$, $c_s^2\equiv P'/\rho'$ and $P\Gamma\equiv
\delta p-c_s^2\delta\rho$.

Finally $\Theta_{\mu\nu}$ is the energy-momentum tensor of the
topological defects. We suppose that it is a small perturbation which
does not contribute to the background. We decompose its scalar
components as~:
\be 
\Theta_{00}=\rho^s,\quad \Theta_{0i}=-D_i v^s,\quad
\Theta_{ij}=\gamma_{ij}\left(P^s-{1\over3}\Delta\Pi^s\right)+
D_iD_j\Pi^s. \label{thetamunu} 
\end{equation}
The four source functions $\rho^s, P^s, v^s, \Pi^s$ will be discussed
later.

The energy and momentum conservation equations for $\Theta_{\mu\nu}$
then read~:
\be
\rho^{s'}+{\cal H}(\rho^s+3 P^s)+\Delta v^s=0 \label{consTD1}
\end{equation}
\be  
v^{s'}+2{\cal H}  v^s+P^s+{2\over3}\Delta \Pi^s=0.\label{consTD2}
\end{equation} 
The conservation equations for the fluid are, at linear order
(see e.g. \cite{kodama84})~: 
\be
\delta'+\frac{1}{2}(1+\omega)h'=-(1+\omega)\Delta v \label{continuity}
\end{equation}
\be
 v'+{\cal H}(1-3c_s^2)v=-\frac{c_s^2}{1+\omega}\delta. \label{euler}
\end{equation}
As for the linearized Einstein equations they can be cast under the form 
(see e.g. \cite{kodama84} or \cite{deruelle98})~:
\be 
-\left(\frac{1}{3}\Delta+K\right)h^-+{\cal H}h'=\kappa( a^2\rho\delta+\rho^s)
\label{einstein1}
\end{equation}
\be 
\left(\frac{1}{3}\Delta +K\right)h^{-\prime}-Kh'=\kappa \left[
a^2\rho(1+\omega)\Delta v+\Delta v^s\right]
\label{einstein2}
\end{equation}
\be
 h''+{\cal H}h'=-\kappa\left[a^2\rho(1+3c_s^2)\delta+(\rho^s+3P^s)\right]
\label{einstein3}
\end{equation}
\be
 (h-h^-)''+2{\cal H}(h-h^-)'-\frac{1}{3}\Delta h^-= -2\kappa \Delta\Pi^s
\label{einstein4}
\end{equation}
where $\kappa\equiv 8\pi G$, $G$ being Newton's constant, and where the
perturbation $h^-$ is defined as
\be 
h^-\equiv h-2\Delta E. \label{defhmoins}
\end{equation}

Equations (\ref{consTD1}-\ref{einstein4}) are eight equations for the
four unknowns $h, h^- $ (or $E$), $\delta$ and $v$~: the source functions 
$\rho^s, P^s, v^s, \Pi^s$, subject to the constraints
(\ref{consTD1}-\ref{consTD2}) being known, equations
(\ref{continuity}-\ref{einstein4}) must give $h, h^-$ (or $E$),
$\delta$ and $v$, two of these equations being redundant. The four Einstein
equations for scalar perturbations (\ref{einstein1}-\ref{einstein4}) include
two constraints on the four variables so that the general solution depends on
two physical degrees of freedom which must be fixed by the physics of the
problem.

To be complete, let us recall the Friedmann equations for the
background
\be
\kappa a^2\rho=3\left({\cal H}^2+K\right),\quad 
\kappa a^2\left(\rho+3P\right)=-6{\cal H}'.\label{friedmann}
\end{equation}

%---------------------------------------------------------------------
\section{Synchronous gauges inside out}
%---------------------------------------------------------------------

\subsection{A toy model~: coherent defects in a flat, radiation dominated, 
universe}\label{toy}

To show, on a simple example, how the above system of equations can be
solved, we consider here a radiation dominated universe with flat,
$K=0$, spatial sections. Hence, $\omega=c^2_s=1/3$, $\Gamma=0$ and,
from (\ref{friedmann}), $a(\eta)\propto \eta$.  Moreover we suppose
that we are deep enough in the radiation era for the perturbations of
interest today to be all larger than the comoving Hubble radius
($1/{\cal H}$). Finally we assume that, in that large scale (or long
wavelength) limit, the variables describing the defects take the form
\be
\rho^s=\eta^{-1/2}A_1,\quad P^s=\eta^{-1/2}A_2,\quad
v^s=-\eta^{1/2}A_3,\quad \Pi^s=\eta^{3/2}A_4\label{td}
\end{equation} 
where $A_i$ are four functions (or, rather, four random variables) of
space varying on scales larger than the Hubble radius. Provided that
the defect network is coherent, uncorrelated on scales larger than the
Hubble radius and scales with the background (the precise definition
of these properties is given in e.g.
\cite{deruelle98}), it can be modelled by (\ref{td}), (see e.g. 
\cite{turok96}, \cite{durrer97}). Now, the precise
energy-momentum tensor of the network of topological defects depends
on the kind of defects considered and can be obtained only by heavy
numerical simulations (see e.g. \cite{durrer96}). It turns out that,
generically, the network is not perfectly coherent so that
$\Delta\Pi^s$, and not $\Pi^s$ itself, varies on scales larger than the
Hubble radius. However such incoherent sources can, in principle, be
decomposed into a sum of coherent sources described by (\ref{td})
\cite{turok97}.

The conservation equations (\ref{consTD1}-\ref{consTD2}) first impose
that, in the long wavelength limit (where we neglect the gradient terms)
\be 
A_1=-6A_2\quad{\rm and}\quad A_3=\frac{2}{5}A_2.\label{tdcon}
\end{equation}
Then equation (\ref{continuity}) and (\ref{einstein3}) can be
rewritten, in the same approximation, using (\ref{td}) and
(\ref{tdcon}), as an evolution equation for $\delta$,
\be
\delta''+\frac{1}{\eta}\delta'-\frac{4}{\eta^2}\delta=-2\kappa A_2\eta^{-1/2}, 
\label{evodelta}\end{equation}
the general solution of which is
\be
\delta=\lambda_0 \eta^2+\frac{4e_0}{\eta^2}+\frac{8}{7}\kappa
A_2\eta^{3/2},
\label{soldelta1}\end{equation}
where $\lambda_0$ and $e_0$ are two constants of integration (that is
two slowly varying functions of space). It is usual to call the
$\lambda_0$ term the ``growing" mode \cite{turok96}. Then the continuity
equation (\ref{continuity}) yields
\be
h=-\frac{3}{2}\lambda_0\eta^2-6\frac{e_0}{\eta^2}-\frac{12}{7}\kappa
A_2\eta^{3/2}-9\Psi_0,\label{solh1}
\end{equation}
$\Psi_0$ being a new function of integration.
The Euler equation (\ref{euler}) then gives $v$,
\be
v=-\frac{\lambda_0}{12}\eta^3+\frac{e_0}{\eta}-\frac{4}{35}\kappa
A_2\eta^{5/2}+\lambda_1 \label{solv1},
\end{equation}
$\lambda_1$ being another function of integration.

At this stage of the resolution we are left with the problem of determining 
the second scalar perturbation of the metric (either $h^-$ or $E$)
with one of the three remaining equations (\ref{einstein1}),
(\ref{einstein2}), (\ref{einstein4}). For instance, the resolution of
(\ref{einstein4}) yields 
\be
E= -e_1- e_0 \ln{\eta}+ \frac{1}{80}\lambda_0\eta^4
-\frac{\Psi_0}{4}\eta^2 -\frac{\Psi_1}{\eta}
+\frac{4}{441}\kappa\left(7A_4-2A_2\right)\eta^{7/2}, \label{solhS1}
\end{equation}
$e_1$ and $\Psi_1$ being yet two other integration functions. 

On the six integration functions that have been introduced, namely
$\lambda_0, e_0, \Psi_0, \lambda_1, e_1,$ and $ \Psi_1$, two of them
($e_0$ and $e_1$) can be eliminated by a gauge transformation (see
eq. \ref{jauge2}). This means that the two remaining Einstein
equations (\ref{einstein1}) and (\ref{einstein2}) are not identically
satisfied, but, rather, should give two constraints on the remaining
four integration constants, leaving out the two physical degrees of
freedom.  This task requires however some shrewdness since the
gradient terms in (\ref{einstein1}) and (\ref{einstein2}) cannot be
simply disposed of.

In \cite{pen94} another route was therefore pursued. First
a ``conserved pseudo-tensor'' $\tau_{\mu\nu}$ defined
 by \cite{shoba90}
\be
\kappa\tau_{00}\equiv\kappa\left(\delta T_{00}+\Theta_{00}\right)-
{\cal H}h',\quad
\kappa\tau_{0k}\equiv\kappa(\delta T_{0k}+\Theta_{0k}),\quad
\kappa\tau_{kl}\equiv\kappa(\delta T_{kl}+\Theta_{kl})-{\cal H}
(h'_{kl}-h'\delta_{kl})
\label{pseudotens}
\end{equation}
was introduced. This quantity is indeed conserved, in that
$\tau_{0i}'=\partial_j\tau_{ij}$ and $\tau_{00}'=\partial_j\tau_{0j}$
provided Einstein's equations (8-17) are satisfied. It has then been
widely used (see e.g. \cite{pen94}, \cite{turok96}, \cite{hu96},
\cite{hu97}) to fix the initial conditions, that is to determine the
value of some of the arbitrary functions of space previously
introduced. The line of the argument is that since the defect network
is uncorrelated on scales larger than the Hubble radius, $\tau_{kl}$
must be white noise on large scales, that is that its components, like
the source functions $A_i$, must be functions of space varying on
scales larger than the Hubble radius. Now, from the conservation laws
one gets that $\tau''_{00}=\partial_{ij}\tau_{ij}$. Hence, if
$\tau_{00}$ and $\tau'_{00}$ are continous through the phase
transition, then $\tau_{00}$ is second order in the gradients, i.e.
\be
\tau_{00}=0\label{tau001}
\end{equation}
on large scales.
 Now, from the
definition (\ref{pseudotens}) and the expressions (\ref{deltaT}),
(\ref{thetamunu}) for $\delta T_{\mu\nu}$ and $\Theta_{\mu\nu}$, we
have
\be
\tau_{00}=\frac{3}{\eta^2}\delta-\frac{6A_2}{\sqrt\eta}-\frac{h'}{\eta}=
6\lambda_0\label{hyp1}
\end{equation}
(using the solution (\ref{soldelta1}) and (\ref{solh1}) for
$\delta$ and $h$). Therefore the integration function $\lambda_0$ is
second order in the gradients and must be set equal to zero. Thus the
use of the ``conserved pseudo-tensor"  (25) eliminates the ``growing" mode of
the density constrast $\delta$.

As for the other integration functions $\Psi_0$, $\Psi_1$ and
$\lambda_1$ entering the solutions (\ref{solh1}), (\ref{solv1}) and (24) for
$h$, $v$ and $E$ ($e_0$ and $e_1$ just reflect the non-unicity of
synchronous gauges) they are usually \cite{pen94} chosen by assuming
 continuity through the phase transition, that is
set equal to zero.

\subsection{Matching conditions in synchronous gauges}

The choices made above to fix the functions of integration (that is to
impose the continuity of $\tau_{00}$, $h$ and $v$ and $E$) must be
justified. The question is therefore to determine which quantities are
indeed continuous through the phase transition. This has been studied
in detail in
\cite{deruelle98}. For perturbations varying on  scales larger
 than the Hubble radius the
phase transition which is at their origin looks instantaneous and the matching
conditions between the pre- and post- transition eras are the standard
ones~: the induced three-metric on, and the extrinsic curvature of the
surface of transition, which is taken to be a surface of constant
density, must be continuous.

In synchronous gauges, these matching conditions translate as
\cite{deruelle98} (we give them for all $K$)
\begin{eqnarray}
\left[h^-+\frac{2}{3({\cal H}^2+K)(1+\omega)}
\left(\kappa a^2\rho\delta+\kappa \rho^s\right) \right]_\pm=0 \label{M1}\\
\left[{\cal H}E'+
\frac{1}{9({\cal H}^2+K)(1+\omega)} 
\left(\kappa a^2\rho\delta+\kappa \rho^s\right)\right]_\pm=0 \label{M2}\\
\left[{\cal H}h^{-\prime}-\left(1-\frac{2}{3}
\frac{K}{({\cal H}^2+K)(1+\omega)}\right)
\left(\kappa a^2\rho\delta+\kappa \rho^s\right)
\right]_\pm=0\label{M3}
\end{eqnarray}
where $[F]_{\pm}\equiv
lim_{\epsilon\rightarrow0^+}\left(F(\eta_{PT}+\epsilon)
-F(\eta_{PT}-\epsilon)\right)$,  $\eta_{PT}$ being the conformal time
at which the transition occurs.  Two useful linear
combinations will turn out to be
\begin{eqnarray}
\left[\left(1-
\frac{2}{9}\frac{1}{({\cal H}^2+K)(1+\omega)}\Delta\right)\left(\kappa
a^2\rho\delta+\kappa
\rho^s\right)-{\cal H}h'+Kh^-\right]_\pm=0 \label{M4}\\
\left[h^--6{\cal H}E'\right]_\pm=0. \label{M5}
\end{eqnarray}

\subsection{Pseudo-tensor and matching conditions}

Let us now study the implications of these general matching
conditions when applied to our toy model. 

First, equation (\ref{M4}), when particularized to a radiation
background, large scale fluctuations and flat spatial sections, is
nothing else than the condition of continuity of $\tau_{00}$.  This
justifies the assumption (\ref{tau001}) and hence the choice
$\lambda_0=0$ (see (\ref{hyp1})).

In order now to assess the validity of the other choices made (continuity
of $h$, $v$ and $E$), we shall turn to the newtonian gauge.  Indeed,
(\ref{M5}) implies, in our toy model, that 
\be
\left[h^--\frac{6}{\eta}E'\right]_\pm=0
\end{equation}
and thus that, see (\ref{defhmoins})
\be
h^-={\cal O}(E),\qquad h={\cal O}(h^-).
\end{equation}
A natural set of variables to solve Einstein's equations (8-15) is
then $(h^-,E)$ or $(h,E)$, but not ($(h-h^-), h$). Now, it also makes
sense to find a combination of the perturbations variables such that
the set of equations (\ref{continuity}-\ref{einstein4}) reduces to a
master equation and to five algebraic equations, two of which being
truly redundant, and not, as is the case in synchronous
gauges, constraints on integration constants. Such a combination is
known and amounts to working in the newtonian gauge
\cite{bardeen80}. Indeed, defining
\be 
\Psi\equiv-\frac{h^-}{6}+{\cal H}E',\quad \Phi\equiv-E''-{\cal H}E',
\quad \delta^\sharp\equiv\delta+3{\cal H}(1+\omega)E',\quad
v^\sharp \equiv v+E',\label{defgi}
\end{equation}
(with ${\cal H}=1/\eta$ and $\omega=1/3$ for our toy model)
it is easy to show that (\ref{continuity}-\ref{einstein4}) reduce to (see e.g.
\cite{deruelle98})~:
\be
\delta^{\sharp '}=-\frac{4}{3}\Delta v^\sharp+4\Psi' \label{eqcon}
\end{equation}
\be
v^{\sharp\prime}+\Phi+\frac{1}{4}\delta^\sharp=0. \label{eqeu}
\end{equation}
\be
\Psi-\Phi=\kappa\Pi^s \label{eqphi}
\end{equation}
\be
\Delta\Psi=\frac{3}{2\eta^2}\left(\delta^\sharp-\frac{4}{\eta}v^\sharp\right)
+\frac{\kappa}{2}\left(\rho^s-\frac{3}{\eta}v^s\right)\label{eqdel}
\end{equation}
\be
\Psi'+\frac{1}{\eta}\Phi=-\frac{2}{\eta^2}v^\sharp-\frac{\kappa}{2}v^s
\label{eqv}
\end{equation}
\be
\Psi''+\frac{4}{\eta}\Psi'-\frac{1}{3}\Delta\Psi=\kappa
\left(\frac{1}{3}\Delta\Pi^s+
\frac{\Pi^{s'}}{\eta} +\frac{1}{2}P^s-\frac{1}{6}\rho^s\right).\label{eqpsi}
\end{equation}
The sources being known, the integration of (\ref{eqpsi}) gives $\Psi$
in function of two integration constants. Then $\Phi$, $v^\sharp$, and
$\delta^\sharp$ follow algebraically from (\ref{eqphi}), (\ref{eqv}),
and (\ref{eqdel}), that is without introducing new constants. More
precisely the solution of this system is (see \cite{deruelle98})
\be
\Psi=\Psi_0+\frac{\Psi_1}{\eta^3}+\frac{2}{9}\kappa\eta^{3/2}(A_4+A_2),\quad
\Phi=\Psi_0+\frac{\Psi_1}{\eta^3}+\frac{1}{9}\kappa\eta^{3/2}(2A_2-7A_4),
\label{solpsi2}
\end{equation}
\be
\delta^\sharp=-2\Psi_0+4\frac{\Psi_1}{\eta^3}+
\frac{8}{9}\kappa\eta^{3/2}(A_2+A_4),\quad
v^\sharp=-\frac{1}{2}\eta \Psi_0 +\frac{\Psi_1}{\eta^2}
-\frac{2}{45}\kappa\eta^{5/2}(4A_2-5A_4).
\label{soldelta2}
\end{equation}
Only two integration ``constants", $\Psi_0$ and $\Psi_1$, are
introduced (and not six as is the case when working in synchronous
gauges), so that the conservation equations (\ref{eqcon}) and
(\ref{eqeu}) are identically satisfied. Note that in the newtonian
gauge, the ``growing" mode is the term proportional to $\Psi_0$ (and
is {\it not} the same as the synchronous growing mode which is
proportional to $\lambda_0$, see (\ref{soldelta1})).

Now the relations (\ref{defgi}) can be inverted to compute
$h, h^-, \delta$ and $v$. Since one has to solve two differential
equations to do so, two functions of space  appear, which can be
identified with $e_0$ and $e_1$, and the comparison of  (\ref{solpsi2}) and
(\ref{soldelta2}) to  (\ref{soldelta1}-\ref{solhS1}) tells us that
\be
\lambda_1=0\quad{\rm and}\quad \lambda_0=0,
\end{equation}
at lowest order in the gradients, which justifies the assumption of
continuity of $v$, and, again, $\tau_{00}$ ($\tau_{00}=0$ for large
scales being nothing more than a rewriting of (\ref{eqdel})).

Finally, the two constants $\Psi_0$ and $\Psi_1$ correspond to the two
physical degrees of freedom and are determined by the matching
conditions. They turn out not to be zero, but the terms proportional to them
in (42), (43)  become nevertheless  quickly negligible
(see
\cite{deruelle98}). Hence  the assumptions that $h$ and $E$ be continuous
are therefore asymptotically valid.

To conclude, working in synchronous gauges introduces additional
integration functions which render the resolution of the perturbation
equations slightly cumbersome. A solution to get rid of these spurious
functions is to go to the newtonian gauge where the perturbation
variables are such that the set of equations
(\ref{continuity}-\ref{einstein4}) reduces to a unique master equation
and algebraic equations, two of which are redundant if the others are
satisfied. The matching conditions at the surface of transition then
fixes the two physical functions and justify the use of $\tau_{00}$ to
eliminate the synchronous ``growing mode'' (i.e. $\lambda_0$), and the
assumptions of continuity of $v$ and (approximately) $h$ and $E$. Now,
no gauge is perfect~: when the sources are incoherent, $\Pi^s$ in the
right-hand-side of (\ref{eqpsi}) is a non local term.

%---------------------------------------------------------------------
\section{Generalization to universes with hyperbolic sections}
%---------------------------------------------------------------------

We now turn to universes with hyperbolic
hypersurfaces ($K=-1$). We will first work in the newtonian
gauge and then generalize the concept of ``pseudo-tensor''.

When the spatial sections are hyperbolic, the curvature generates a new
length scale on top of the comoving Hubble or horizon scale
($R_H=1/{\cal H}$), which is the comoving curvature scale ($R_C=1$). As
follows from (\ref{friedmann}), the curvature scale is larger than
the horizon scale
\be
R_H<R_C.
\end{equation}
We are interested in all scales $L$ that were larger than the horizon
scale at the phase transition. They fall in three categories~: (I)
those which entered the horizon between the phase transition and today
(i.e. $L\leq R_{H_0}$), (II) those which have not yet entered the
horizon but are subcurvature (i.e. $R_{H_0}\leq L\leq R_C$), (III)
those which are supercurvature (i.e.  $L>R_C$).

\subsection{Newtonian gauge}

As we have seen, the introduction of the newtonian gauge perturbations
(\ref{defgi}) was useful to keep to a minimum the number of integration
functions. The  generalization to $K\neq0$ spacetimes governed by a
mixture of fluids of the master equation for
$\Psi$ (\ref{eqpsi}) is (see e.g.
\cite{kodama84})
\begin{equation}
\Psi''+3(1+c_s^2){\cal H}\Psi'+[2{\cal H}'+({\cal H}^2-K)(1+3c_s^2)]\Psi
-c_s^2\Delta\Psi={\cal S} \label{EVO}
\end{equation}
where ${\cal S}$ is a source involving the variables describing the
topological defects, whether they satisfy (\ref{tdcon}) or not.
Equation (\ref{EVO}) is closed, in the sense that ${\cal S}$ is a
known function of time, only in the case when the background is
governed by a single fluid.  In the general case of a multi-fluid
system, one has to add a set of equations for the relative entropies
which enter ${\cal S}$ (see
\cite{kodama84}). However, on super-horizon scales, all these entropies
turn out to be constant so that the multi-fluid can be described as a single
one and the equation is again closed. In this paragraph, we therefore restrict
our attention to modes which have not yet entered the horizon that
is such that ($L>R_{H_0}$), i.e. from categories II and III.
For those modes, it is easy to show that the term $c_s^2\Delta\Psi$
is negligible compared to $({\h}^2-K)\Psi$ in the regime when ${\h}^2$
is of order $K$ (see Appendix 1).

Equation (\ref{EVO}) can then be rewritten under quite a
compact form as
\begin{equation}
\chi'=\frac{{\cal H}}{{\cal H}^2-{\cal H}'+K}
{\cal S},\label{CHIPRIME}
\end{equation}
where we have introduced instead of $\Psi$ the new variable $\chi$ defined
by
\begin{eqnarray}
\chi&\equiv&\frac{{\cal H}^2}{{\cal H}^2-{\cal H}'+K}\frac{1}{a^2}
\left(\frac{\Psi a^2}{{\cal H}}\right)'.\label{DEFCHI}
\end{eqnarray}
The general solution of (\ref{CHIPRIME}) can be written formally  as
\begin{equation}
\chi=\chi_0+\int^\eta\!d\bar\eta\,\frac{{\cal H}}{{\cal
H}^2-{\cal H'}+K}{\cal S}\label{CHI}
\end{equation}
where $\chi_0$ is an integration constant, that is a slowly varying function
of space only. Inverting equation (\ref{DEFCHI}), we 
can therefore write the general solution of equation (\ref{EVO}) on
superhorizon scales as
\begin{equation}
\frac{{a^2}}{\cal H}\Psi=\chi_1-\chi_0\int^\eta \!\!d\bar\eta\, a^2 
\frac{{\cal H}^2-{\cal H}'+K}{{\cal H}^2}-\int^\eta \!\!d\bar\eta a^2 
\frac{{\cal H}^2-{\cal H}'+K}{{\cal H}^2}
\int^{\bar\eta}\!\!d\tilde\eta\,\frac{{\cal H}}{{\cal
H}^2-{\cal H'}+K}{\cal S}\label{SOLPSI}
\end{equation}
where $\chi_1$ is another integration function. $\chi_0$ and $\chi_1$ are
determined by the  matching conditions as in the flat case (in which case
they are proportional to $\Psi_0$ and $\Psi_1$).

(Note that in the absence of the source term ${\cal S}$, as is the
case in inflationary models, $\chi$ generalizes Lyth's
variable $\zeta$ \cite{lyth} to Friedmann-Lema\^\i tre spacetimes with
hyperbolic spatial sections. Its constancy on superhorizon scales
(equation (\ref{CHI})) can then be used to compute, by means of
(\ref{SOLPSI}), the amplification of $\Psi$ in open inflationary
models.)

As for the other perturbations, they are given algebraically by $$
\Phi=\Psi-\kappa\Pi^s,\quad v^\sharp=-\frac{2}{3({\cal
H}^2+K)(1+\omega)}\left(\Psi'+{\cal H}\Phi+ \frac{\kappa}{2}
v^s\right),$$ \be \delta^\sharp=\frac{1}{3({\cal
H}^2+K)}\left\lbrace\kappa(\rho^2-3{\cal H}v^s) -6{\cal H}(\Psi'+{\cal
H}\Phi)\right)\rbrace.  \end{equation} (which generalize equations (\ref{eqphi})
(\ref{eqdel}) (\ref{eqv}) (\ref{eqpsi})). The solution depends only on
the two integration functions $\chi_0$ and $\chi_1$, and the continuity
and the Euler equations are identically satisfied.

\subsection{Synchronous gauges}

Provided we have solved the equations of the background
(\ref{friedmann}), the set of equations (\ref{continuity},
\ref{euler}, \ref{einstein3}) can be solved as in the section
\ref{toy}.  However, the question is then to find a prescription to
fix the coefficient of the synchronous ``growing mode'' of the density
contrast (i.e. the equivalent of the constant $\lambda_0$ in
(\ref{soldelta1})). For that purpose, we will generalize $\tau_{00}$.

If we note that, when $K=0$, $\tau_{00}=0$, $\tau_{00}$ being given by
(\ref{pseudotens}), is in fact a way to write the matching condition
(\ref{M4}) at lowest order, a straightforward generalization of
$\tau_{00}$ is
\be
\kappa\tau_{00}\equiv\sqrt{\gamma}\left\lbrace\kappa\delta T_{00}+
\kappa\Theta_{00}+Kh^- -{\cal H}h'\right\rbrace\label{defti0}
\end{equation}
where $\gamma$ is the determinant of the 3-metric
$\gamma_{ij}$. Indeed, if we introduce
\be
\kappa\tau_{0k}\equiv\sqrt{\gamma}\left\lbrace\kappa\delta T_{0k}
-2K\partial_kE'\right\rbrace,\label{defitilde}
\end{equation}
it can easily be checked, using the
Einstein equations (\ref{einstein1}, \ref{einstein2}), that $\tau_{0\nu}$
is conserved, in the sense that
\be
\partial_0 \tau_{00}=\partial_k \tau_{0k}.\label{PRD54}
\end{equation}
Now, the matching condition (\ref{M4}) imposes that initially
\be
\tau_{00}=0\label{tau002}
\end{equation}
at lowest order on superhorizon scales. Furthermore (\ref{PRD54}) implies
that $\tau_{00}$ remains zero at all times for superhorizon modes. Thus
(\ref{tau002}) is a way to eliminate at all times the spurious growing mode in
the solution for the density contrast $\delta$.

%---------------------------------------------------------------------
\section{The matching conditions in terms of a vector conservation
law}
%---------------------------------------------------------------------

In the previous section we have encoded part of the matching
conditions into a quantity $\tau_{00}$, the geometrical status of
which is not defined. We show here that for long wavelength modes,
it can be interpreted in terms of a conserved vector
related to a symmetry of the background. For that purpose, we briefly
describe the formalism we use and then apply it to our problem.

\subsection{General Formalism}

A general formalism was developped by 
Katz-Bi${\rm \check c}$\'ak-Lynden Bell \cite{katz97} to
define conserved quantities and conservation laws in an arbitrary
spacetime (${\cal M}$,$g_{\mu\nu}$). This generalizes the work
by Bergmann who defined conservation laws with respect to a 
flat background \cite{bergmann49}. The formalism involves the 
introduction of a background ($\bar {\cal M}$,${\bar g}_{\mu\nu}$)
and a mapping, i.e. a way to identify points of
$\cal M$ and $\bar{\cal M}$.

The central idea is that one can construct a lagrangian
for the gravitational field, quadratic in the
covariant derivatives of the metrics and normalized so that it
vanishes on the background (i.e. when ${\cal M}={\bar {\cal M}}$), which is a
true scalar density. From this quantity, one can therefore build
vectors densities
${{\hat I}^\mu}$, which are   conserved in the sense that
${\partial_\mu}{{\hat I}^\mu}=0$.

The detailed construction of the vector densities ${{\hat I}^\mu}$ and
the demonstration of their properties can be found in
\cite{katz97}. We will just sum up the
relations which will be useful and fix the notations. The
conserved vector field density ${{\hat I}^\mu}$ is given by
\begin{equation}
\kappa{{\hat I}^\mu}\equiv \left[\left({\hat
G}^\mu_\nu - {\bar {\hat G}^\mu_\nu}\right) + \frac{1}{2}{\hat
l}^{\rho\sigma}{\bar R}_{\rho\sigma}\delta^\mu_\nu +\kappa{\hat
t}^\mu_\nu\right]\xi^\nu +\kappa{\hat\sigma}^{\mu[\rho\sigma]}
\partial_{[\rho}{ \xi}_{\sigma]} + \kappa{\hat
{\cal Z}}^\mu(\xi^\nu), \label{DEFI}
\end{equation}
where $\xi^\mu$ is an arbitrary vector field. 
For any quantity $A$, ${\hat A}$ denotes $\sqrt{-g} A$, $g$ being the determinant of
the metric $g_{\mu\nu}$,  and ${\bar A}$ the value of $A$ on the background. 
${\hat G}^\mu_\nu$ and ${\bar
{\hat G}^\mu_\nu}$ are the Einstein tensor densities of the spacetime and of
the background,  ${\hat l}^{\mu\nu}$ is the
difference of the two metric tensor densities,
$${\hat l}^{\mu\nu} \equiv {\hat g}^{\mu\nu} -{\bar {{\hat g}} }
^{\mu\nu},$$ 
${\bar R}_{\mu\nu}$ is the Ricci tensor of the
background spacetime. Introducing 
$${\Delta}^\lambda_{\mu\nu} \equiv
\Gamma^\lambda_{\mu\nu} - {\bar \Gamma}^\lambda_{\mu\nu},$$
where 
$\Gamma^\lambda_{\mu\nu}$ and ${\bar \Gamma}^\lambda_{\mu\nu}$ are
the Christoffel symbols of the spacetime and of the background, as well as
${\bar\nabla}_\mu$ and
$\nabla_\mu$ the two covariant derivatives associated with 
$\bar{g}_{\mu\nu}$ and $g_{\mu\nu}$,
the expressions for ${\hat t}^\mu_\nu$, ${\hat\sigma}^{\mu[\rho\sigma]}$ 
and ${\hat {\cal Z}}^\mu (\xi^\nu)$ are 
\begin{eqnarray}
 2\kappa
  {\hat t}^\mu_\nu & = &{\hat g}^{\rho\sigma}
  \left({\Delta}^\lambda_{\rho\lambda}{\Delta}^\mu_{\sigma\nu} +
  {\Delta}^\mu_{\rho\sigma}{\Delta}^\lambda_{\lambda\nu} -
  2{\Delta}^\mu_{\rho\lambda}{\Delta}^\lambda_{\sigma\nu}\right) \nonumber \\
&&+ {\hat g}^{\mu\rho}\left({\Delta}^\sigma_{ \lambda\sigma}
{\Delta}^\lambda_{\rho\nu} -
   {\Delta}^\sigma_{\sigma\rho}{\Delta}^\lambda_{\lambda\nu}\right)
-{\hat g}^{\rho\sigma}
   \left({\Delta}^\eta_{\rho\sigma}{\Delta}^\lambda_{\lambda\eta} -
  {\Delta}^\eta_{\rho\lambda}
  {\Delta}^\lambda_{\eta\sigma}\right)\delta^\mu_\nu 
\end{eqnarray}
(this term reduces
to the Einstein pseudo-tensor density when the background is
Minkowski spacetime in cartesian coordinates),
\begin{equation}2\kappa {\hat
\sigma}^{\mu[\rho\sigma]} =\left({\hat l}^{\mu[\rho} 
{\bar g}^{\sigma]\lambda} -
{\bar g}^{\mu[\rho}{\hat l}^{\sigma]\lambda}\right)
{\Delta}^\nu_{\lambda\nu} -2 {\hat l}^{\lambda[\rho}{\bar
g}^{\sigma]\nu} {\Delta}^\mu_{\lambda\nu}
\end{equation}
(where $[]$ means antisymmetrization), and
\begin{eqnarray}
4\kappa {\hat {\cal Z}}^\mu (\xi^\nu) &=& \left(Z^\mu_\rho 
{\hat g}^{\rho\sigma} +
  {\hat g}^{\mu\rho}Z^\sigma_\rho
  -{\hat g}^{\mu\sigma}Z\right){\Delta}^\lambda_{\sigma\lambda} \nonumber \\
&&+\left({\hat g}^{\rho\sigma}Z 
-2{\hat g}^{\rho\lambda}Z^\sigma_\lambda\right)
  {\Delta}^\mu_{\rho\sigma} +{\hat g}^{\mu\lambda}\partial_\lambda Z   
+{\hat g}^{\rho\sigma}\left({\bar\nabla}^\mu Z_{\rho\sigma}-2{\bar\nabla}_\rho
    Z^\mu_\sigma\right),\label{DEFCALZ} 
\end{eqnarray}
where $Z_{\rho\sigma}$ and $Z$ are given by
\begin{equation}Z_{\rho\sigma} = {\cal L}_\xi
{\bar g}_{\rho\sigma} = 2{\bar\nabla}_{(\rho} \xi_{\sigma)}\;\;  
{\rm and} \;\;  Z= Z_{\rho\sigma} {\bar g}^{\rho\sigma}
\label{DEFZ}\end{equation}
(where () means symmetrization).

Let us emphasize that the relation 
\begin{equation}
\partial_\mu {\hat I}^\mu=0\label{conslaw}
\end{equation}
 is an
identity which holds for all ($g_{\mu\nu}$, ${\bar g}_{\mu\nu}$, $\xi^\mu$).
The vector field $\xi^\mu$ can be seen
as generating an infinitesimal
change of coordinates~: contrarily to special relativity, one can build
an infinite numbers of conserved vectors corresponding
to all reparametrizations.
These conserved vectors are useful to write integral
conservation laws (known as ``strong conservation laws"). When 
$\xi^\mu$ is a Killing vector field of the background, these
strong conservation laws are called ``Noether conservation laws", i.e.
conservation laws related to a symmetry. 

We also stress that up to now, we have done no linearization; the case
when ${\cal M}$ is a ``small perturbation" of ${\bar{\cal M}}$ is
considered in the next section.

\subsection{Conserved vectors in a
perturbed Friedmann-Lema\^\i tre spacetime}

We now apply this general formalism to perturbations in cosmology.
The manifold ${\cal M}$ is a perturbed Friedmann-Lema\^\i tre universe
and the background is taken to be a Friedmann-Lema\^\i tre spacetime, as it
is usual in the theory of cosmological perturbations. Let us note that other
choices for the background can be made, for instance de Sitter spacetime
(see e.g.
\cite{deruelle97}). The line element of ${\cal M}$ can then be written as
\begin{equation}
ds^2\equiv g_{\mu\nu}dx^\mu dx^\nu=\left(\bar{g}_{\mu\nu} + h_{\mu\nu}
\right)dx^\mu dx^\nu,
\end{equation}
where $h_{\mu\nu}$ is a ``small perturbation".

We can now linearize equation (\ref{DEFI}). A first step yields \cite{katz97}
\begin{equation}
\kappa{\hat I}^\mu=\sqrt{-\bar g}\left[\delta G^\mu_\nu+
\frac{1}{2}\left(\bar R^\mu_\nu\delta^\sigma_\rho -
\bar R^\sigma_\rho\delta^\mu_\nu\right)h^\rho_\sigma\right]\xi^\nu 
+\kappa{\hat{\cal Z}}^\mu+{\cal O}(h^2).
\end{equation}

The second step is to linearize ${\cal Z}^\mu$. To do that we need to
specify the arbitrary vector field $\xi^\mu$. As discussed before the
relation $\partial_\mu{\hat I}^\mu=0$ is valid whatever field we
use. However Killing vectors of the background (such that
${\bar\nabla}_{(\mu} \xi_{\nu)}=0$) are of special interest since they
are related to symmetries and have a geometrical interpretation.
Unfortunately the six Killing vectors of the Friedmann-Lema\^\i tre
background lay in spacelike hypersurfaces and cannot generate any quantity
related to time translation, that is to energy.  Now, we know that when two
spacetimes are conformal, the Killing vectors of one of the spacetimes
are at least conformal Killing vectors (that is such that
${\bar\nabla}_{(\mu} \xi_{\nu)}=\frac{1}{4}\bar
g_{\mu\nu}\bar\nabla_\rho\xi^\rho$) of the other \cite{eisenhardt50}.
We also know that the Friedmann-Lema\^\i tre spacetimes are conformal to
Einstein, Minkowski and de Sitter spacetimes \cite{hawking73}, each of
which has a Killing vector field associated with time translation that
induces a conformal Killing vector field on the Friedmann-Lema\^\i tre
spacetimes.  In the following, we shall use one particular conformal Killing
vector to generalize the symmetry under time translation. Let us first
construct this vector field explicitely.

The background metric being written as
\begin{equation}
d\bar{s}^2\equiv\bar{g}_{\mu\nu}dx^\mu dx^\nu=a^2(\eta)\left[-d\eta^2+
\gamma_{ij}dx^idx^j\right]\quad\hbox{with}\quad
\gamma_{ij}=
\delta_{ij}+K\frac{\delta_{im}\delta_{jn}x^mx^n}{1-K\delta_{mn}
x^mx^n}\label{METRIC}
\end{equation}
where $\gamma_{ij}$ is the spatial metric in Weinberg's
coordinates \cite{wein},  we  choose 
the conformal Killing vector
defined by
\begin{equation}
\xi^\mu=\delta^\mu_0, \label{xi}
\end{equation}
which is a Killing vector of 
Einstein static spacetimes, i.e. obtained by setting $a=1$ in
(\ref{METRIC}). This vector is associated with translations
in cosmic time, i.e. in the proper time
of  a comoving observer of the Friedmann-Lema\^\i tre background.  Other
conformal Killing vectors could have been used (for instance the one
associated with the conformity to de Sitter spacetime, see e.g.
\cite{deruelle97}), but (\ref{xi}) is the useful one for the law we want to
generalize.

The vector $\xi^\mu$ having thus being chosen, we can now linearize
${\cal Z}^\mu$. We see from equation (\ref{DEFZ}) that 
${\cal Z}^\mu$ vanishes for any Killing vector field. 
For any conformal Killing vector field 
equation (\ref{DEFCALZ}) gives after linearization \cite{katz97},
\begin{equation}
8\kappa{\cal Z}^\mu=\left(h \bar g^{\mu\rho}-h^{\mu\rho}\right)\partial_\rho Z-
Z\bar\nabla_\rho\left(h \bar g^{\mu\rho}-h^{\mu\rho}\right),
\end{equation}
with $h=h_{\mu\nu} \bar g^{\mu\nu}$ and  $Z$ given
by equation (\ref{DEFZ}).

Let us now, as a third step,  specialize to scalar perturbations, that is
to the sub-class of perturbations in synchronous gauges as defined in 
(\ref{metric}).
We now have all the elements to compute the conserved vector associated to
the conformal Killing
vector field (\ref{xi}). One can first easily show that
\be Z_{\mu\nu}=2{\cal H}g_{\mu\nu}\qquad {\rm and}\qquad Z=8{\cal H},
\end{equation}
and a straightforward calculation then yields~:
\begin{eqnarray}
\kappa{\hat I}^0 &=& a^4\sqrt{\gamma}\left[\kappa\delta T^0_0+\kappa\Theta^0_0
-\frac{K}{a^2} h + \frac{{\cal H}}{a^2}{h'}\right]
\nonumber \\
\kappa{\hat I}^k &=&a^4\sqrt{\gamma}\left[\kappa\delta T^k_0+\kappa\Theta^k_0+
\frac{{\cal H}}{a^2} (D^lh^k_l-D^kh)\right],\label{defihat}.
\end{eqnarray} 
In synchronous gauge (\ref{defihat}) can be written in terms of $E$ and $h$ as
\begin{eqnarray}
\kappa{\hat I}^0 &=& -a^2\sqrt{\gamma}\left[\kappa\left(\delta T_{00}+
\Theta_{00}\right)+Kh-{\cal H}h'\right]\nonumber\\
\kappa{\hat I}^k &=&a^2\sqrt{\gamma}\left[\kappa \left(\delta T_{0k}+
\Theta_{0k}\right)+2{\cal H}\gamma^{kl}\partial_l\left(E-\frac{1}{3}h^-\right)
\right],\label{defihat2}
\end{eqnarray} 
It is clear that the identity (\ref{conslaw}) with $\hat I^\mu$ given
by (\ref{defihat}) or (\ref{defihat2}) is a consequence 
of the Einstein constraint equations
(\ref{einstein1}-\ref{einstein2}), but a useful one as we shall see.

\subsection{Encoding the matching conditions in a conserved vector field}

Comparing the expression of $\tau_{00}$ (\ref{defti0}) and $\tau_{0k}$ 
(\ref{defitilde}) to
$\hat I^\mu$ (\ref{defihat2}), and recalling that $h=h^-$ on
superhorizon scales, we have that, in that limit~:
\be
\tau_{00}=-\frac{\hat I^0}{a^2},\quad
\partial_k\tau_{0k}=\frac{1}{a^2}\left[\partial_k\hat I^k+2{\cal H}\hat I^0
\right].
\end{equation}
Thus, for superhorizon scales, the matching condition (\ref{M4}) can be
encoded in the components of a conserved vector field related to a
symmetry since 
\be
\left[\tau_{00}\right]_\pm=0\qquad\Longleftrightarrow\qquad
\left[\hat I^0\right]_\pm=0.
\end{equation}
This enables us
to relate $\tau_{\mu0}$, in the long wavelength limit, to a vector field
built with the conformal Killing vector field $\xi^\mu$ associated
with time translation.  In fact the construction of $\tau_{\mu\nu}$
uses in an obvious way the fact that a Friedmann-Lema\^\i tre universe
with flat spatial sections is conformal to Minkowski spacetime (see
Verraraghavan and Stebbins
\cite{shoba90} section VI and Pen et al. \cite{pen94} section IV-F).

$\tau_{00}$ is often referred to as a ``pseudo-energy''
\cite{turok96}.  Our construction can justify such a denomination
 since it is related to a symmetry involving time translation.

%---------------------------------------------------------------------
\section*{Aknowledgments} JPU and ND thanks J. Katz, 
D. Langlois, P. Peter and A. Riazuelo for many enlightening
discussions and the programme Alliance who enabled them to visit
DAMTP. We are grateful to our referee whose remarks greatly helped to
improve the presentation of this paper.
%--------------------------------------------------------------------- 

\section*{Appendix 1}

We compare the terms $c_s^2\Delta\Psi$ and $({\h}^2-K)(1+3c_s^2)\Psi$
in equation (\ref{EVO})
for modes of characteristic length $L$ (i.e. such that $\Delta\Psi\sim
\Psi/L^2$)
in a universe with hyperbolic spatial sections ($K=-1$) and dominated
by a mixture of radiation (subscript $r$) and dust (subscript $m$).

Using the fact that $P=P_r$ and $\rho=\rho_m+\rho_r$, we can relate
the sound speed $c_s^2\equiv P'/\rho'$ to the energy density of the radiation
and the dust by
\begin{equation}
c_s^2=\frac{1}{3}\frac{1}{1+\frac{3}{4}\frac{\rho_m}{\rho_r}}.
\end{equation}
If we now introduce $x\equiv a/a_0$ and $\Omega_0^X\equiv
\kappa\rho_0^X/3H_0^2$ (where a subscript $0$ means that we evaluate the
quantity today and $X$ stands for either $r$ or $m$),
 and use the Friedmann equation
(\ref{friedmann}), we have that
\begin{equation}
c_s^2\Delta\Psi\ll({\h}^2-K)(1+3c_s^2)\psi\Longleftrightarrow
L^{-2}\ll 3\left(1+\frac{3}{4}x\frac{\Omega_0^m}{\Omega_0^r}\right)
\left(2+\frac{\Omega_0^m}{1-\Omega_0}\frac{1}{x^2}\left[x+
\frac{\Omega_0^r}{\Omega_0^m}\right]\right).\label{ineq1}
\end{equation}
All modes belonging to the categories (II) and (III) are such that
\begin{equation}
L^{-1}<R_{H_0}^{-1}.
\end{equation}
Therefore, the inequality (\ref{ineq1}) is satisfied for all $x$ such that
\begin{equation}
1+\frac{\Omega^r_0}{1-\Omega_0}\ll3\left(1+\frac{3}{4}x\frac{\Omega_0^m}
{\Omega_0^r}\right)
\left(2+\frac{\Omega_0^m}{1-\Omega_0}\frac{1}{x^2}\left[x+
\frac{\Omega_0^r}{\Omega_0^m}\right]\right).\label{ineq2}
\end{equation}
For $\Omega_0^m\sim 0.2$, 
it is easy to see that this inequality is satisfied for all $x$ in
$]0,1]$.

\end{document}